# Observation of a node in the quantum oscillations induced by microwave radiation

Alexey E. Kovalev,[1] Sergey A. Zvyagin,[2] Clifford R. Bowers,[1]
John L. Reno[3] and Jerry A. Simmons[3]

[1]Department of Chemistry, University of Florida, Gainesville, FL, 32601

[2]National High Magnetic Field Lab, Tallahassee, FL 32310

[3]Sandia National Laboratories, MS 1415, Albuquerque, NM 87185

## ABSTRACT

The microwave induced magnetoresistance in a GaAs/AlGaAs heterostructure was studied at temperatures below 1K and frequencies in the range of 150-400 GHz. A distinct node in the Shubnikov –de Haas oscillations, induced by the microwave radiation, is clearly observed. The node position coincides with the position of the cyclotron resonance on the carriers with effective mass $(0.068 \pm 0.005)\, m_0$.

The discovery of the microwave induced magnetoresistance oscillations (MIMO), and microwave induced zero resistance states in 2D electron systems [1-4] has stimulated much activity in this area. Recently these states were studied in the Corbino ring geometry [5]. However, despite the numerous theoretical papers [6-16] proposing explanations of these phenomena, there are still some problems in understanding the effect.

The MIMO is analogous to the Shubnikov–de Haas oscillations (SdHO) in that both are periodic functions of the inverse magnetic field, but the periodicity of the former is determined by the ratio of microwave frequency $\omega$ to the cyclotron frequency $\omega_c$, while the SdHO period is determined by the cross-sectional area of the Fermi surface. As a consequence of the small cyclotron mass (~0.065 $m_0$, where $m_0$ is the free electron mass) and relatively high two dimensional electron densities of $>2\times10^{11}$ cm$^{-2}$, the MIMO and zero resistance states were observed predominantly at fields below the onset of the SdHO. The purpose of the present work is to expand the range of MIMO studies to low filling factors by employing a high power, high frequency microwave source. Such data may provide new facts for verification of theories and lead to a better understanding of the phenomenon.

While previous studies [1-4] in structures with mobilities of up to $25\times10^6$ cm$^2$/Vs have shown that the MIMO phenomenon is most strongly developed in samples with the highest mobility, the present work is restricted to sample of moderate mobility (~$5\times10^6$ cm$^2$/Vs, $1.9\times10^{11}$ cm$^{-2}$ density) due to their immediate availability. The 2DES was located at GaAs/Al$_x$Ga$_{1-x}$As (x=0.3) heterojunction. The doping was provided by a $\delta$-doped Si layer setback from the heterojunction. A standard Hall bar pattern was used to measure the longitudinal magnetoresistance ($R_{xx}$). The sample exhibits MIMO at frequencies below 60 GHz at a temperature of about 0.4 K (not shown). The magnetoresistance measurements were performed using the standard 4-point contact technique at an *ac* frequency of 77 Hz and a current of 1 μA. The sample was placed at a distance of 1 cm from the end of a stainless circular waveguide with a 10 mm inner diameter, and immersed in liquid $^3$He. The measurement temperature was approximately 0.5-0.7 K, depending on the radiation power. The experiments were performed using high-field mm- and submm- wave spectroscopy facilities [17] at the National High Magnetic Field Laboratory in Tallahassee. The microwave radiation, which was produced by a set of backward wave oscillators operating in the 150-400 GHz range, is assumed to be unpolarized at the output of the waveguide.

By pushing the MIMO to higher fields we effectively went to a regime where both MIMO and SdHO can be observed simultaneously, as shown in Figure 1. Here, two curves are plotted, one acquired with and one without microwave irradiation at 285 GHz. In addition to the MIMO, seen below 0.5 T, we found quite a new phenomenon: a node in the SdHO induced by the microwave radiation. For the radiation frequency of 285 GHz the node position $B_{node}$ was 0.7 T. We note that the SdHO phase remains the same on both sides of $B_{node}$. As is apparent in Figure 1, both curves have minima at the same magnetic field. This suggests that if the SdHO amplitude is modulated by some function, which is zero at $B_{node}$, that this function does not change sign at the node.

The node was also observed at several other frequencies. In Figure 2 we plot the node position versus the radiation frequency, ν. The solid line represents a linear least squares fit of $B_{node}(ν)$, demonstrating that $B_{node}$ is proportional to ν. From the slope of the fit, we conclude that the node position coincides with the position of the cyclotron resonance of the carriers with an effective mass $(0.068 \pm 0.005)\ m_0$.

In Figure 3 we plot the microwave induced change in the magnetoresistance versus inverse field. We see that the MIMO maxima are equally spaced, but the distance from $B_{node}$ to the nearest maximum is significantly less than the field independent spacing between maxima. This means that if $B_{node}$ corresponds the cyclotron resonance, the positions of the MIMO maxima would be described by

$$\omega/\omega_c = n-\varphi, \; n=2,3,4... \tag{1}$$

This formula is closer to the one reported by Mani et al [3] then to the one reported in Ref. [1-2, 16].

To illustrate the presence of the phase shift we plotted the positions of the maxima together with the node positions scaled by the radiation frequency, as shown in Figure 4. The nodes occur at index 1, while the maxima begin at index 2. We see that all the maxima lie on a straight line, which has an offset of about 0.3 at $\nu/B=0$. This is close to the offset value of 0.25 reported in Ref. [3].

Finally, we should mention that the value of the cyclotron frequency obtained by fitting of the MIMO maxima to Equation 1 is within experimental error the same as the one deduced from the frequency dependence of $B_{node}$.

The presence of the node means that the change in conductivity due to the microwave radiation at the cyclotron resonance is not simply an additive contribution. If that were the case the SdHO could be interpreted as the interference between the dark conductivity and the photoconductivity, and the amplitude of the SdHO would not vanish, even if the SdHO from photoconductivity would have a node. From our data it is clearly apparent that the SdHO are modulated by microwaves. This suggests that MIMO is not a photoconductivity effect, but is instead the result of some change in electron scattering under the microwave influence. Indeed, considering that the total microwave power in such experiments is on the order of several milliwatts, and that the typical area of the microwave field is about 1 cm$^2$, one can estimate the electric field to be about 1 V/cm. The average electron velocity in the microwave field in the high mobility samples will be comparable to the Fermi velocity. That means that the conductivity response to the *dc* and microwave electric fields cannot be separated.

We cannot say anything definite about the possible existence of other nodes in the SdHO for other integer values of $\omega/\omega_c$ because the oscillation amplitude is too small. It would be very interesting to investigate this in a sample with higher mobility where higher index nodes might be observed.

In conclusion, we have studied the microwave induced magnetoresistance oscillations in a region where they overlap with the Shubnikov–de Haas oscillations. In a sample with moderately high 2D electron mobility, we found a clear node in the Shubnikov–de Haas oscillations that is induced by the microwave radiation. The node position coincides with the position of the cyclotron resonance with effective mass $0.068m_0$. In accordance with a previous study, the maxima of the microwave induced magnetoresistance oscillations are shifted from the positions of the cyclotron resonance harmonics.

**FIGURE CAPTIONS**

**Figure 1.** Longitudinal magnetoresistance acquired with (solid line) and without (dashed line) microwave radiation. A clear node is apparent at about 0.7 T. Note that the microwave radiation does not change the phase of the Shubnikov-de Haas oscillations. The measurement temperature was about 0.6 K.

**Figure 2.** Plot of node position versus radiation frequency. A least squares fit to the data yields an effective mass of $(0.068 \pm 0.005)m_0$.

**Figure 3.** Microwave induced component of the magnetoresistance, $\Delta R_{xx}$, obtained during irradiation at a frequency of 285 GHz, versus inverse static magnetic field. Note that the frequency difference between the node and the first maximum is smaller then the frequency separation of the maxima.

**Figure 4.** Maxima and nodes obtained with different microwave frequencies. Nodes are marked by index 1, while maxima are numbered starting with index 2.

**ACKNOWLEDGEMENTS**

This work was supported by NSF grant DMR-0106058 and the University of Florida. A portion of this work was conducted at the National High Magnetic Field Laboratory. The NHMFL is supported by NSF Cooperative Agreement No. 0084173 and by the State of Florida.

**LITERATURE**

1. M. A. Zudov, R. R. Du, J. A. Simmons, and J. L. Reno, cond-mat/9711149; Phys. Rev. B **64**, 201311(R) (2001).
2. P. D. Ye, L. W. Engel, D. C. Tsui, J. A. Simmons, J. R. Wendt, G. A. Vawter, and J. L. Reno, Appl. Phys. Lett. **79**, 2193 (2001).
3. R. G. Mani, J. H. Smet, K. von Klitzing, V. Narayanamurti, W. B. Johnson, and V. Umansky, Nature (London) **420**, 646 (2002).
4. M. A. Zudov, R. R. Du, L. N. Pfeiffer, and K. W. West, Phys. Rev. Lett. **90**, 046807 (2003).
5. C. L. Yang, M. A. Zudov, T. A. Knuuttila, R. R. Du, L. N. Pfeiffer, and K. W. West, Phys. Rev. Lett. **91**, 096803 (2003).
6. J. C. Phillips, cond-mat/0212416.
7. A. C. Durst, S. Sachdev, N. Read, and S. M. Girvin, Phys. Rev. Lett. **91**, 086803 (2003).
8. A. V. Andreev, I. L. Aleiner, and A. J. Millis, Phys. Rev. Lett. **91**, 056803 (2003).
9. P. W. Anderson and W. F. Brinkman, cond-mat/0302129.
10. J. Shi and X. C. Xie, Phys. Rev. Lett. **91**, 086801 (2003).

11. A. A. Koulakov and M. E. Raikh, cond-mat/0302465.

12. F. S. Bergeret, B. Huckestein, and A. F. Volkov, Phys. Rev. B **67**, 241303(R) (2003).

13. I. A. Dmitriev, A. D. Mirlin, and D. G. Polyakov, Phys. Rev. Lett. **91**, 226802 (2003).

14. I.A.Dmitriev, M.G.Vavilov, I.L.Aleiner, A.D.Mirlin, D.G.Polyakov, cond-mat/0310668

15. S. A. Mikhailov, cond-mat/0303130.

16. M. A. Zudov, Phys. Rev. B 69 (2004) 041304.

17. S.A. Zvyagin, J. Krzystek, P.H.M. van Loosdrecht, G. Dhalenne, and A. Revcolevschi, Phys. Rev. B 67 (2003) 212404.

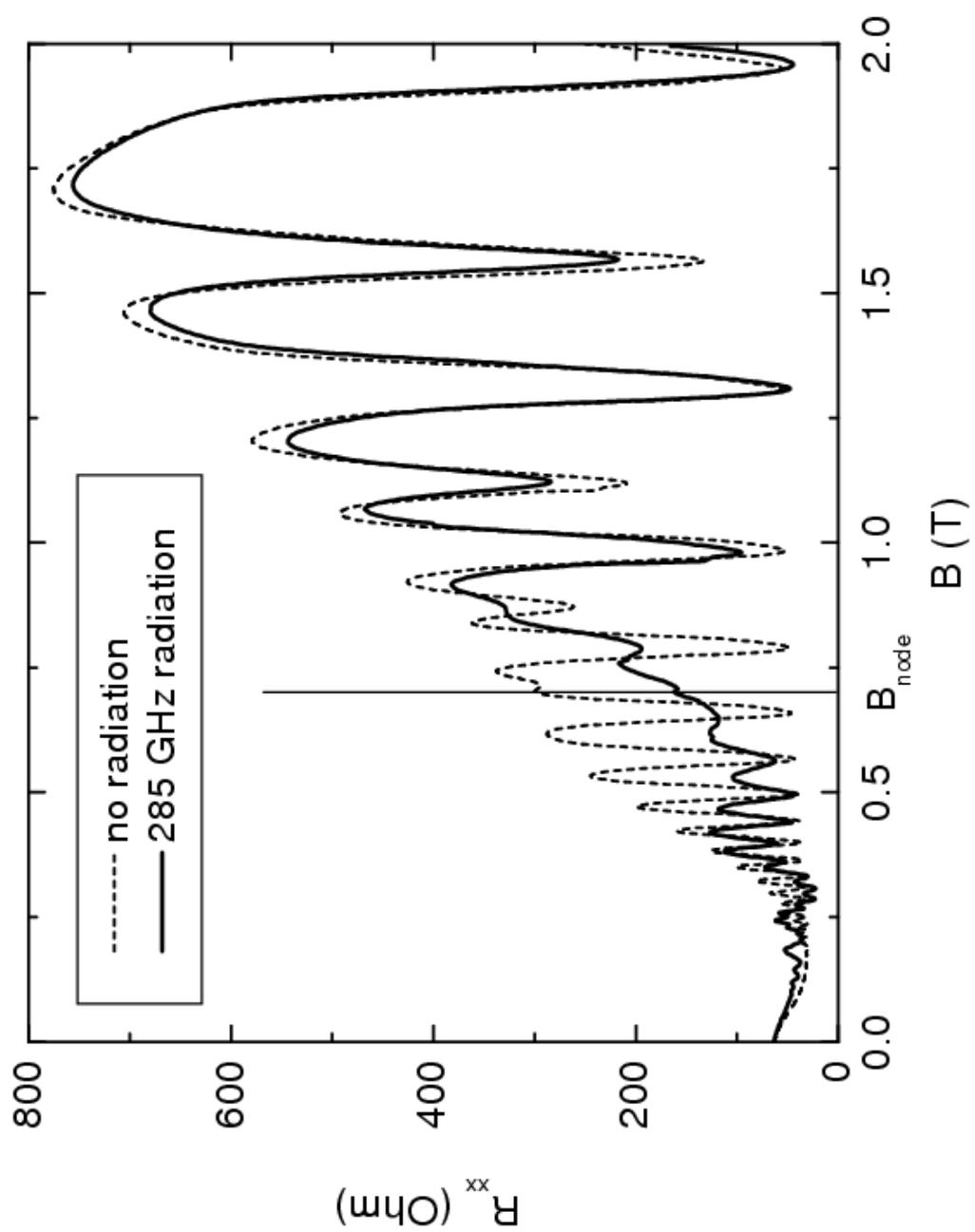

Fig 1.

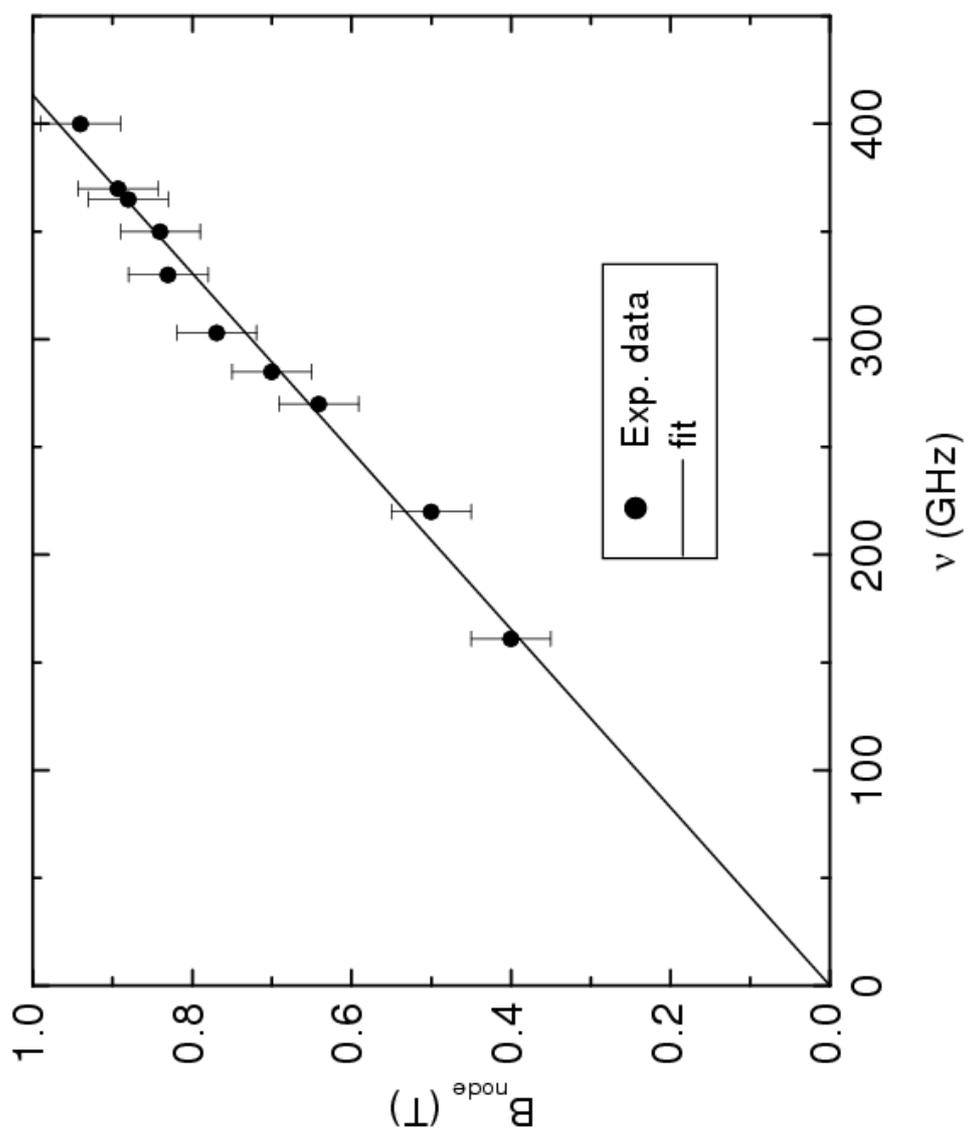

Fig. 2

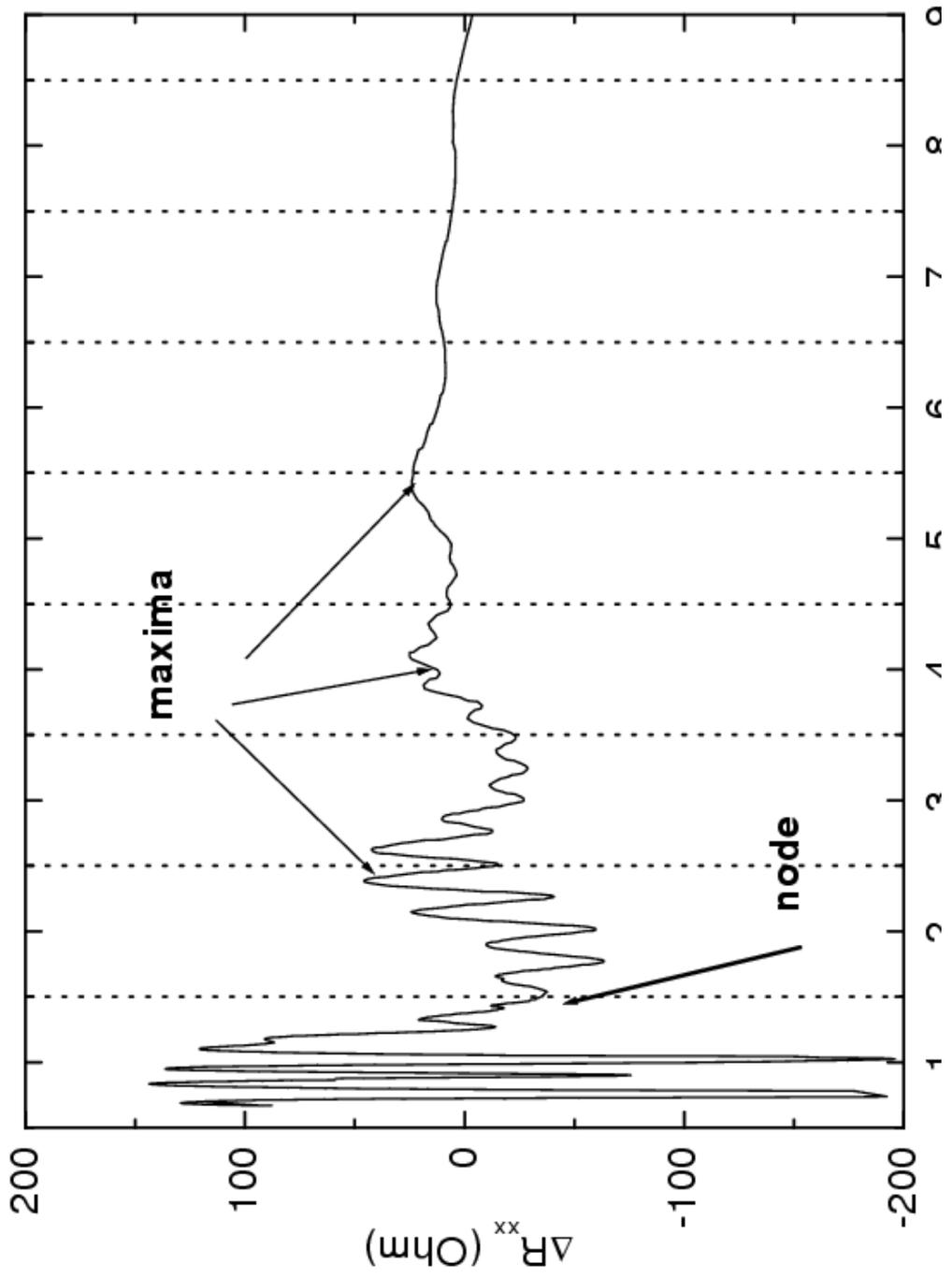

Fig.3

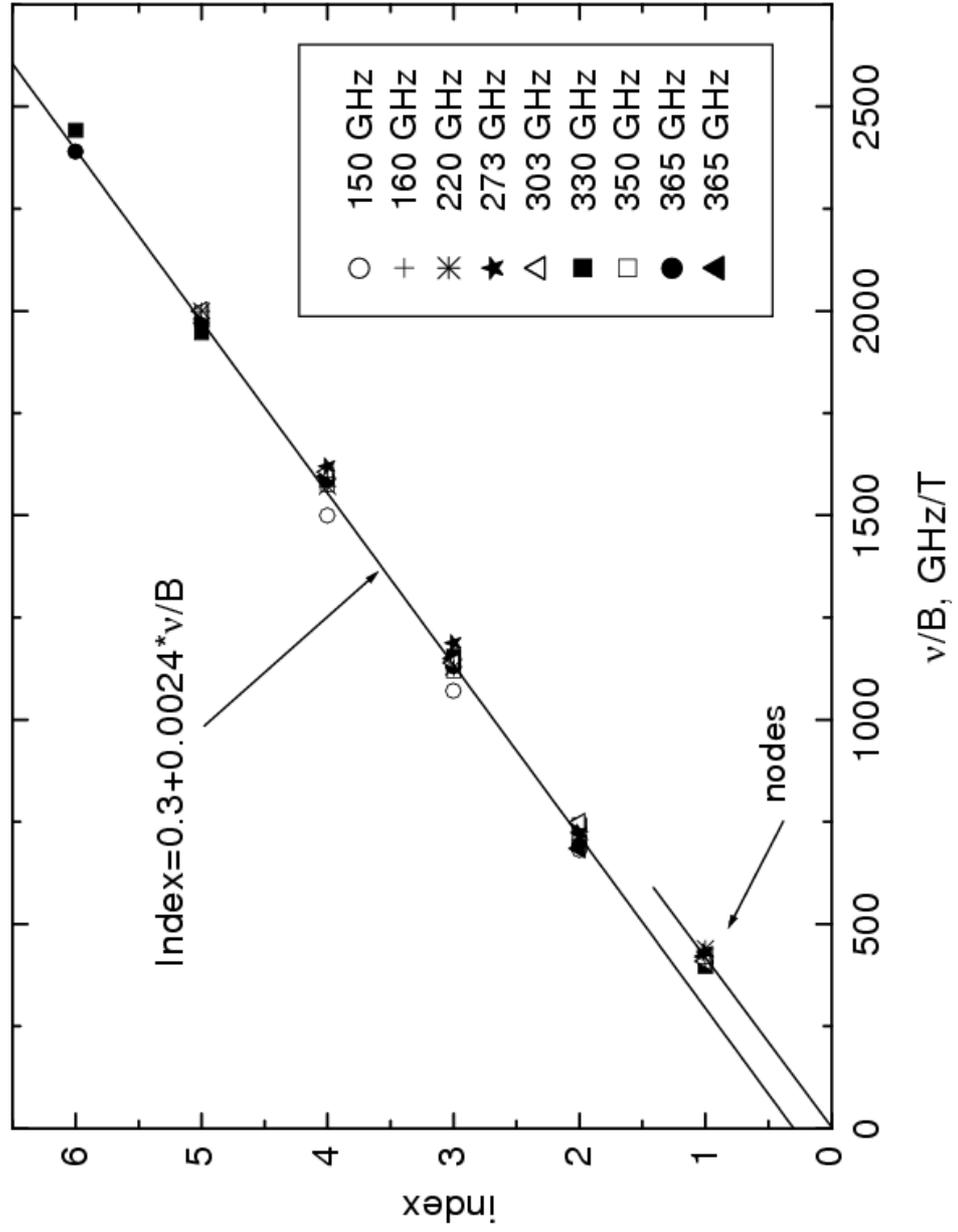

Fig. 4